\documentclass[aps,prl,twocolumn,showpacs,superscriptaddress,amssymb]{revtex4}
\usepackage{graphicx}
\usepackage{hyperref}
\usepackage{natbib}
\usepackage{bm}
\begin{document}

\title{Bootstrap tomography of high-precision pulses for quantum control}

\author{V. V. Dobrovitski}
\affiliation{Ames Laboratory US DOE, Iowa State University, Ames IA 50011, USA}
\author{G. de Lange}
\affiliation{Kavli Institute of Nanoscience Delft, Delft University of Technology, 
P.O. Box 5046, 2600 GA Delft, The Netherlands}
\author{D. Rist\`e}
\affiliation{Kavli Institute of Nanoscience Delft, Delft University of Technology, 
P.O. Box 5046, 2600 GA Delft, The Netherlands}
\author{R. Hanson}
\affiliation{Kavli Institute of Nanoscience Delft, Delft University of Technology, 
P.O. Box 5046, 2600 GA Delft, The Netherlands}

\date{\today}

\begin{abstract}
Long-time dynamical decoupling and quantum control of qubits
require high-precision control pulses. Full characterization
(quantum tomography) of imperfect pulses presents a bootstrap problem:
tomography requires initial states of a qubit which 
can not be prepared without imperfect pulses. We present a
protocol for pulse error analysis, specifically tailored for
a wide range of the single solid-state electron spins.
Using a single electron spin of a nitrogen-vacancy (NV)
center in diamond, we experimentally verify the correctness of
the protocol, and demonstrate its usefulness for 
quantum control tasks.
\end{abstract}


\maketitle

Coherent manipulation of single and few electron spins has recently 
been achieved in several solid-state systems such as quantum 
dots and diamond defect centers. 
Such systems are promising candidates for quantum
information processing~\cite{LD98,qrepeater}, precise metrology
\cite{qmetrology} and ultra-sensitive magnetometry \cite{qmagnetNV}.
They also present an excellent testbed for studying the
fundamental problems of quantum dynamics of open systems 
\cite{HansonNature08,openquantum,HamEng}.
High-speed manipulation of the system's quantum state 
can be achieved
by using microwave or optical pulses~\cite{fastrotSemi,fastrotNV2,fastrotNV}, 
which must be fine-tuned to provide a high degree of fidelity. 
For example, sequences of quantum control pulses can be used to
extend the coherence time via dynamical decoupling
~\cite{ddViola,ddCDD,ddions,ddmolecules,fullerenes}. 
For long sequences, even small errors in the pulses will destroy 
the coherence that one attempts to
preserve~\cite{ddmolecules,errHans} and may even lead to artificial
saturation~\cite{freez1,freez2}. Therefore, precise characterization 
of errors is essential for successful implementation of complex 
quantum control protocols. With known errors, composite pulses
and/or special pulse sequences can be chosen to mitigate the problem.

Complete information on the action of a pulse can in
principle be gained with quantum process tomography (QPT)
\cite{NielsenChuangBook}. However, QPT of an imperfect
pulse requires preparation and measurement of a complete set
of reference states, whereas in many solid-state qubit
systems (e.g.\ quantum dots, diamond defect centers,
superconducting circuits) only one state can be prepared
reliably (without the imperfect pulses), and only one
observable can be directly measured. All other states can be
prepared only with the imperfect pulses themselves, and
therefore have errors~\cite{ddJJ1}. This presents a
bootstrap problem: the reference states contain the very
same errors that we want to determine.

The problem of pulse error analysis has been studied
extensively in the areas of NMR and
ESR~\cite{GerDyb,BurumEtal81,ShakaEtal88,Haub79}. However, single
electron spins in solid-state settings present new opportunities and
challenges, and call
for new approaches tailored at the specific demands of these systems. 
The driving pulse field can be tightly confined in the 
vicinity of the target spin.
The resulting strong, nanosecond-timescale pulses enable fast 
spin manipulation, but 
the standard pulse error analysis~\cite{GerDyb,BurumEtal81,ShakaEtal88,Haub79}
used in NMR becomes inapplicable. 
At strong driving, the spin dynamics changes noticeably \cite{fastrotNV}.
The non-secular terms in the rotating frame
can become important. The ac-Stark and Bloch-Siegert 
shifts can significantly detune the pulse
frequency from resonance~\cite{fastrotNV} and tilt the rotation axis
towards the $z$-axis. Also, the pulse edges 
constitute
a much larger fraction of the short pulse, and the driving field 
at the edges varies much faster and 
stronger than in typical NMR pulses. The resulting
errors~\cite{fastrotNV} (e.g.\ tilting of
the rotation axis) can go beyond
the standard treatment~\cite{decohDuringPulses},
and can not always be removed by symmetrizing the pulse shape.

Also, typical NMR systems have long coherence times that exceed the
pulse width by orders of magnitude. The standard
tune-up protocols~\cite{GerDyb,BurumEtal81,ShakaEtal88,Haub79} 
exploit this advantage, and use sequences with tens or hundreds of
pulses to achieve outstanding precision in pulse parameters. 
But single solid-state electron spins are dephased faster, on a
timescale $T_2^*$ of microseconds down to tens of
nanoseconds~\cite{HansonNature08}. After only tens of pulses
the signal becomes a complex mixture of pulse errors and decoherence
\cite{freez1,freez2}. To ensure a reliable measurement of the
errors, the protocol for single electron spins must be short, a few
pulses at most.

Here, we present a systematic approach to pulse characterization for
single solid-state electron spins, which is 
usable at shorter coherence times
and much stronger driving power compared to traditional NMR systems.
The proposed protocol contains four series of measurements, each
having only 1--3 pulses, thus minimizing the effect of
decoherence. The measured signal quantifying the pulse errors grows
linearly with the errors 
to ensure a good accuracy for small errors. Also, the
signal is zero for zero errors for good relative accuracy.
The protocol determines all pulse errors: the rotation angle and all
three components of the rotation axis \cite{decohDuringPulses}.
We experimentally
demonstrate the protocol on a single spin of a nitrogen-vacancy (NV)
defect center in diamond. By deliberately introducing known pulse
errors, we verify the accuracy and self-consistency of
the protocol, and use it to significantly increase the fidelity of
QPT. 

Our goal is to determine the parameters of
four pulses, $\pi_X$, $\pi_Y$, $\pi/2_X$, and $\pi/2_Y$ 
applied to a two-level quantum system 
($\pi_X$ denotes a rotation by an angle $\pi$ around the $x$-axis 
in the rotating frame; other notations are analogous). 
These pulses allow implementation
of universal decoupling XY sequences~\cite{ddViola,ddCDD},
full tomography of the density matrix, 
and universal single-qubit gates~\cite{NielsenChuangBook}.
We assume that the pulse errors are reasonably small, and consider
only the first-order terms in these quantities (since we want the
signal to grow proportionally to errors).
We also assume that the pulse width $t_p$ is small in
comparison with the dephasing time $T_2^*$;
in this case the impact of decoherence is of second order,
$(t_p/T_2^*)^2$, and is negligible for short sequences \cite{decohDuringPulses}.
Under this assumption the evolution of a spin 
during the pulse can be described a unitary rotation. For example, 
for $S=1/2$,
the evolution (in the rotating frame) during
an imperfect $\pi_X$ pulse is given by
\begin{equation}
\label{eq:error_example}
U_X={\rm e}^{-i(\vec n\vec\sigma)(\pi+2\phi)/2}\approx -\phi-i(\sigma_x%
+\epsilon_y\sigma_y+\epsilon_z\sigma_z),
\end{equation}
where $\sigma_{x,y,z}$ are the Pauli matrices,
the rotation angle error is $2\phi$ and the rotation axis $\vec n$ has 
small components $n_y=\epsilon_y$ and $n_z=\epsilon_z$. Similarly, a $\pi/2_X$ pulse $U'_X$ has the rotation
angle error $2\phi'$, and the small rotation axis components $\epsilon'_y$ and $\epsilon'_z$ along
$y$ and $z$, respectively.
Note that in general two $\pi/2$ pulses do not yield the same evolution as
one $\pi$ pulse due to errors introduced by the pulse edges. Analogous parameters for $y$-pulses will be denoted as $2\chi$, $v_x$, and $v_z$ (angle and axis errors for $\pi_Y$), and $2\chi'$, $v'_x$, and $v'_z$ (angle and axis errors for $\pi/2_Y$).

The bootstrap protocol shares ideas with the standard QPT, 
and with the NMR tune-up sequences. 
Before each measurement, the spin is in
the state $|\uparrow\rangle$, and the measured signal is 
$\langle\psi|\sigma_z|\psi\rangle$,
where $|\psi\rangle$ is the wavefunction after the pulse.
An imperfect pulse $U_j$ can be represented as a product 
$U_j=U_j^{(0)} V_j \approx U_j^{(0)}({\mathbf 1}-i K_j)$, where $U_j^{(0)}$ is
a corresponding ideal rotation and the Hermitian operator $K_j$ is proportional
to small pulse errors. 
Applying two pulses $U_1$ and $U_2$ in succession, we obtain 
up to linear order in $K_j$
\begin{equation}
U_{21} = U_2 U_1 \approx U_2^{(0)} U_1^{(0)} - i U_2^{(0)} K_1 -i K_2 U_1^{(0)},
\end{equation}
and the terms $U_2^{(0)} K_1$ and $K_2 U_1^{(0)}$ contain
different matrix elements of the operators $K_1$ and $K_2$. E.g.,
if $U_1$ and $U_2$ are the (imperfect) $\pi/2_Y$ and $\pi/2_X$ rotations,
the signal detected after this sequence,
$S_{21}={\rm Tr}(\sigma_z U_{21}|\uparrow\rangle\langle\uparrow|U^\dag_{21})$,
contains a linear combination of the
matrix elements $\langle\uparrow|K_1|Y\rangle$
and $\langle\uparrow|K_2|X\rangle$ (where
$|Y\rangle=|\uparrow\rangle+i|\downarrow\rangle$ and 
$|X\rangle=|\uparrow\rangle+|\downarrow\rangle$).
Combining different pulses, we obtain a sufficient number of such linear
combinations of various matrix elements of $K_j$ to uniquely determine all
of them. A general approach to bootstrap tomography can be formulated in
the language of QPT, by expanding the operation element 
operators \cite{NielsenChuangBook} in terms of small errors, and various 
bootstrap protocols applicable to more complex systems can be designed 
in a similar manner. Here, we focus on a protocol for a single two-level system.

\begin{table} [tbp]
\caption{Summary of the bootstrap protocol: pulse sequences (read from
right to left) and the resulting signals expressed in terms of the error parameters. 
Blocks of sequences are separated by horizontal lines.}
\label{table}
\begin{ruledtabular}
\begin{tabular}{cc}
Sequence & Signal \\
$\pi/2_X$ & $-2\phi'$\\
$\pi/2_Y$ & $-2\chi'$\\ \hline
$\pi/2_X$-$\pi_X$ & $2(\phi+\phi')$ \\
$\pi/2_Y$-$\pi_Y$ & $2(\chi+\chi')$\\
$\pi_Y$-$\pi/2_X$ & $-2v_z+2\phi'$\\
$\pi_X$-$\pi/2_Y$ & $2\epsilon_z+2\chi'$\\ \hline
%
$\pi/2_Y$-$\pi/2_X$ & $-\epsilon'_y -\epsilon'_z -v'_x -v'_z 		$\\
$\pi/2_X$-$\pi/2_Y$ & $-\epsilon'_y +\epsilon'_z -v'_x +v'_z 		$\\
$\pi/2_X$-$\pi_X$-$\pi/2_Y$ & $-\epsilon'_y +\epsilon'_z +v'_x -v'_z 	+2\epsilon_y	$ \\
$\pi/2_Y$-$\pi_X$-$\pi/2_X$ & $-\epsilon'_y -\epsilon'_z +v'_x +v'_z 	+2\epsilon_y	$ \\
$\pi/2_X$-$\pi_Y$-$\pi/2_Y$ & $ \epsilon'_y -\epsilon'_z -v'_x +v'_z 	+2v_x				$ \\
$\pi/2_Y$-$\pi_Y$-$\pi/2_X$ & $ \epsilon'_y +\epsilon'_z -v'_x -v'_z 	+2v_x				$ 
\end{tabular}
\end{ruledtabular}
\end{table}

The proposed protocol is summarized in Table~\ref{table}. It consists of three blocks of measurement sequences. For each sequence the measured signal is given in terms of the error parameters. The first block, with two single-pulse sequences, yields the rotation angle errors for the $\pi/2$ pulses. This information is then used in the second block, consisting of four two-pulse sequences, to find the rotation angle errors and the components of the rotation axis along $z$ for the $\pi$ pulses. The third block has six multi-pulse sequences, yielding six signals that are linearly related to the remaining six pulse error parameters. 
This linear system is underdetermined, since the whole system of pulses
is invariant under rotations around the $z$-axis. 
We may put $\epsilon'_y=0$, taking the phase of the $\pi/2_X$ pulse 
as the $x$ direction in the rotating frame. This fixes all other directions,
and all errors are uniquely determined. 
No unphysical results, typical for experimental implementations of 
standard QPT \cite{ddJJ1,qptNV}, appear in this bootstrap protocol. 

We now demonstrate and verify the protocol experimentally by applying
it to a single solid-state spin system. We use the spin of a single
Nitrogen-Vacancy (NV) center, which is a defect in diamond composed of a
substitutional nitrogen atom with an adjacent vacancy~\cite{ddNV1}. 
The NV center's spin can be
optically initialized and read out 
\cite{ddNV1}. The experiments are performed in a home-built confocal
microscope at room temperature. NV centers in nanocrystals are
prepared on a chip with a lithographically-defined waveguide allowing
fast and precise spin rotations by magnetic resonance 
\cite{decohDuringPulses}.

\begin{figure} [tbp]
\includegraphics[width=3.4in]{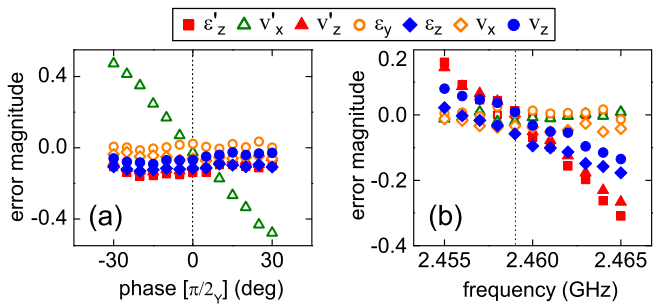}
\caption{(Color online) 
Experimental verification of the bootstrap protocol by introducing 
varying pulse errors. Duration of the $\pi/2$-pulses ($\pi$-pulses) is 5~ns 
(9~ns). \textbf{(a)} Measured error parameters for different phases 
$\Phi$ of the $\pi/2_Y$-pulse. The frequency of the driving field is set at 
2.4605~GHz. 
\textbf{(b)} Measured error parameters for various frequencies of the 
driving field. Error bars everywhere are smaller than the symbol size.
}
\label{fig1}
\end{figure}

We controllably introduce two types of pulse errors, and use the
bootstrap protocol to extract their values. In the first experiment,
we vary the phase $\Phi$ of the nominal $\pi/2_Y$-pulse 
between $-30^\circ$ and
$30^\circ$ from its nominal value. This way, we are
changing the error parameter $v'_x = \sin{\Phi}\approx \Phi{\rm (rad)}$ 
while leaving all other error
parameters constant. Figure~\ref{fig1}(a) shows the results from the
bootstrap protocol that clearly support this expectation.

In the second experiment we detune the microwave excitation away from
the qubit transition frequency, thereby varying the $z$-components of
the rotation axis for all pulses. As shown in Fig.~\ref{fig1}(b), the
extracted error parameters $v_z$, $v'_z$, $\epsilon_z$, and
$\epsilon'_z$ strongly change (roughly linearly) with the detuning as
expected, while the other error parameters stay virtually constant. 
If these rotation axis errors were arising only from the
bulk of the pulse, they would all show the same dependence on the detuning,
and all four curves would have the same slope. Instead we
observe that the errors of the nominal $\pi/2$-pulses vary about twice
as much as the errors of the nominal $\pi$-pulses. This indicates that the
errors originate largely from the pulse edges. 
Since the edges are the same for all pulses, 
they have a relatively larger impact on shorter pulses. The data in
Fig.~\ref{fig1}(a)-(b) demonstrate that the bootstrap protocol is indeed an
efficient and reliable tool for extracting pulse errors.

Due to experimental limitations it may be impossible to cancel all
errors at once. In that case, the choice of the optimal working point
involves a trade-off, and precise knowledge of the pulse errors
becomes particularly important. 
For example, when performing QPT, a set of the reference 
states is prepared using the pulses $\pi_X$, $\pi/2_X$, and $\pi/2_Y$.
These states are acted upon by the process, 
and rotated to the readout basis before measurement
\cite{NielsenChuangBook}. 
The operation elements of the quantumn process 
are expanded in the basis $E_0=I$, $E_1=\sigma_x$, $E_2=\sigma_y$,
and $E_3=\sigma_z$, and the process is completely characterized by
the $4\times 4$ expansion matrix $\chi$~\cite{NielsenChuangBook}.
%
When systematic pulse errors are present, the prepared initial 
states differ from the reference states, and the read-out 
is also performed in 
the incorrect basis, yielding an incorrect matrix $\chi$.
But with pulse errors known, 
the raw measured data can be transformed
into the correct basis prior to the standard QPT data processing
\cite{NielsenChuangBook,ddJJ1,qptNV}.

As a demonstration, and 
as a check of self-consistency of the bootstrap protocol, we perform
QPT while introducing the same pulse errors as in Fig.~1.
We show that with the pulse errors deduced with the protocol, the QPT results can be corrected. 
%
The comparison between raw and corrected data below is
designed to use no {\it a priori\/} assumptions about correctness
of the bootstrap protocol.

First, we take the (imperfect) $\pi_Y$ pulse as
an example of a quantum process. We introduce errors in the QPT procedure by changing the 
phase $\Phi$ of the nominal $\pi/2_Y$-pulse from $-30^\circ$ to $30^\circ$. 
We first determine the reference process
which corresponds to $\Phi=0$;
the corresponding experimental settings and the pulse
error parameters are marked in Figs.~1 and 2 by dotted lines.
We perform QPT on this reference process, and
the resulting reference matrix $\chi_0$ 
is calculated in two ways: (i) using the raw uncorrected data, 
i.e.\ assuming that the pulses used for QPT are ideal (we denote
this matrix as $\chi_0^{\rm r}$), 
and (ii) using the data corrected for the known pulse imperfections (the
resulting matrix is $\chi_0^{\rm c}$). 
Next, we vary $\Phi$, and use artificially deteriorated $\pi/2_Y$ pulses
to determine the matrix $\chi$ of the quantum process. This
matrix is also determined in two way, by using raw
experimental data (matrix $\chi^{\rm r}$), and by correcting 
the data for the known
pulse errors (matrix $\chi^{\rm c}$). 
For each value of $\Phi$, 
we compare the raw-data matrices $\chi^{\rm r}$ and $\chi_0^{\rm r}$ on
one hand, and the corrected 
matrices $\chi^{\rm c}$ and $\chi_0^{\rm c}$ on the other. 

The process we are studying does not
depend on the phase of the nominal $\pi/2_Y$ pulse.
Thus, ideally, the matrices $\chi_0$ and $\chi$ should be the same.
To quantify the difference between $\chi_0$ and $\chi$,
we use two distance measures. One is the process fidelity~\cite{NielsenChuangBook} 
$F=\rm{Tr} [\chi_0 \chi]$, which depends quadratically
on the pulse errors. The other measure is the Hilbert-Schmidt 2-norm 
$||M||_2 = \sqrt{{\rm Tr}[M M^{\dagger}]}$ of the difference matrix 
$M=\chi-\chi_0$. This norm is linear in, and thus more sensitive to, 
the pulse errors \cite{Ref2norm}. 
%
%
%
%
\begin{figure}[tb]
\includegraphics[width=3.4in]{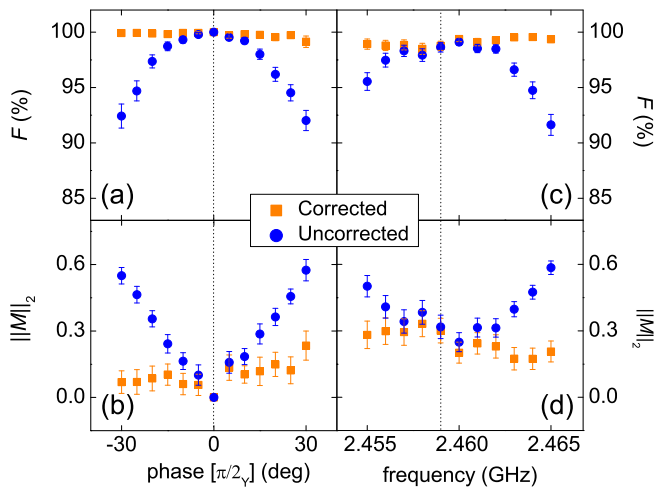}
\caption{(Color online) Correction of pulse errors in Quantum Process Tomography using the bootstrap protocol. \textbf{(a)} Fidelity ($F$) and \textbf{(b)} the 2-norm distance $||M||_2$ between the process measured at finite introduced $\pi/2_Y$ phase error and the process matrix measured at zero introduced error. The process is a $\pi_Y$-pulse with zero introduced error. \textbf{(c)} $F$ and \textbf{(d)} $||M||_2$ between the measured process and the actual process (identity). All measures are calculated both for the uncorrected and for the corrected data.
}
\label{fig2}
\end{figure}

In Figs.~\ref{fig2}(a)-(b), orange squares show the values of $F$ 
and $||M||_2$ 
for the corrected-data matrices $\chi_0^{\rm c}$ and $\chi^{\rm c}$.
The expectation that $\chi_0$ and $\chi$ should coincide
is confirmed with excellent precision. Almost independently
of $\Phi$, the
fidelity remains above 99\%, and $||M||_2$ stays small.
This is not so for the raw-data matrices $\chi_0^{\rm r}$ 
and $\chi^{\rm r}$ (blue squares). 
The neglected phase error of the nominal $\pi/2_Y$ pulse
makes the matrix $\chi^{\rm r}$ inaccurate, so $F$ and $||M||_2$
depend on $\Phi$, with fidelity dropping by 8\%
for $\Phi=30^\circ$.

In a second experiment, we detune the microwave
excitation frequency away from the qubit transition,
introducing the errors $\epsilon_z$, $\epsilon'_z$, $v_z$,
and $v'_z$ (see Fig.~\ref{fig1}b) into all pulses.
The tomographed process is the identity process, which
is independent of the pulse carrier frequency. 
We perform QPT on this process for the same range of
detunings as in Fig.~\ref{fig1}(b). As above, we determine
the corrected and the uncorrected matrices $\chi^{\rm r}$,
and $\chi^{\rm c}$, while the matrix $\chi_0$, describing
the identity process, is independent of the pulse errors.
The results are
shown in Fig.~\ref{fig2}(c)--(d). Again, the fidelities are high
for the corrected data for the full range of introduced
errors, while for the uncorrected data the fidelity has
dropped by as much as 10\%. The same behavior
is seen for $||M||_2$.
Thus, even the effects of complex pulse errors introduced 
by detuning the frequency
can be efficiently corrected using the information
from the bootstrap protocol. 

Summarizing, we have developed an effective pulse error
analysis protocol tailored to the specific requirements of
single solid-state spins. We have experimentally implemented
and verified the protocol on a single electron spin in
diamond. We have shown that the distortions of the
tomography results, arising due to the pulse errors, can be
corrected with the knowledge obtained from the bootstrap
protocol. The methods described in this paper are applicable
to many systems. They may help in accurate determination of the properties of different
quantum processes, a key feature for the fields of quantum
information processing, quantum metrology and fundamental
studies of quantum decoherence.

We would like to thank D. G. Cory, S. Lyon, A. Tyryshkin,  
M. Pruski, C. Ramanathan, K. Schmidt-Rohr, and M. Laforest 
for very useful and enlightening discussions.
Work at Ames Laboratory was supported by the Department of Energy 
--- Basic Energy Sciences under Contract No.~DE-AC02-07CH11358.
We gratefully acknowledge support from FOM, NWO and the DARPA QuEST program.


\begin{thebibliography}{99}

\bibitem{LD98}
D. Loss and D.P. DiVincenzo, Phys. Rev. A 57, 120 (1998).

\bibitem{qrepeater} L. Childress {\it et al.\/}, 
Phys. Rev. Lett. {\bf 96}, 070504 (2006).

\bibitem{qmetrology} J. A. Jones {\it et al.\/}, Science {\bf 324},
1166 (2009);
P. Cappellaro {\it et al.\/},
Phys. Rev. Lett. {\bf 94}, 020502 (2005); 
J. M. Geremia, J. K. Stockton, and H. Mabuchi, Phys. Rev. Lett. {\bf 94}, 
203002 (2005).

\bibitem{qmagnetNV} J. M. Taylor {\it et al.\/}, 
Nature Physics {\bf 4}, 810 (2008); 
C. Degen, Appl. Phys. Lett.~\textbf{92}, 243111 (2008);
G. Balasubramanian {\it et al.\/}, 
Nature {\bf 455}, 648 (2008).

\bibitem{HansonNature08}
R. Hanson and D. D. Awschalom, Nature {\bf 453}, 1043 (2008).

\bibitem{openquantum} L. Childress {\it et al.\/}, Science {\bf 314}, 281 (2006);
R. Hanson {\it et al.\/},
Science {\bf 320}, 5874 (2008); 
D. J. Reilly {\it et al.\/},
Phys. Rev. Lett. {\bf 101}, 236803 (2008),
C. Latta {\it et al.\/},
Nature Physics {\bf 5}, 758 (2009);
I. T. Vink {\it et al.\/}, 
Nature Physics {\bf 5}, 764 (2009).

\bibitem{HamEng} C. H. Tseng {\it et al.\/}, 
Phys. Rev. {\bf A 61}, 012302 (1999); S. Lloyd, Science {\bf 273}, 1073 (1996).

\bibitem{fastrotSemi} J. Berezovsky {\it et al.},
Science {\bf 320}, 349 (2008); 
D. Press {\it et al.\/}, 
Nature {\bf 456}, 218 (2008);
Y. Wu {\it et al.\/}, 
Phys. Rev. Lett. {\bf 99}, 097402 (2007).

\bibitem{fastrotNV2}  P. Neumann {\it et al.\/},
Science {\bf 320}, 1326 (2008); L. Jiang {\it et al.\/},
Science {\bf 326}, 267 (2009).

\bibitem{fastrotNV} G. D. Fuchs {\it et al.\/},
Science {\bf 326}, 1520 (2009).

%

%

\bibitem{ddViola} L. Viola, E. Knill, and S. Lloyd, Phys. Rev. Lett. {\bf 82},
2417 (1999);
L. F. Santos and L. Viola, Phys. Rev. Lett. {\bf 97},
150501 (2006).


\bibitem{ddCDD} K. Khodjasteh and D. Lidar, Phys. Rev. Lett. {\bf 95},
180501 (2005).



\bibitem{ddions} M. J. Biercuk {\it et al.\/}, 
Nature {\bf 458}, 996 (2009).

\bibitem{ddmolecules} J. Du {\it et al.\/}, Nature {\bf 461}, 1265 (2009).
\bibitem{errHans} H. De Raedt {\it et al.\/}, Prog. Theor. Phys.
Suppl. {\bf 145}, 233 (2002).

\bibitem{fullerenes} J. J. L. Morton {\it et al.\/}, 
Nature Phys. {\bf 2}, 40 (2006).

\bibitem{freez1} W. Zhang {\it et al.\/}, 
Phys. Rev. B {\bf 77}, 125336 (2008);
W. Zhang {\it et al.\/},
J. Phys.: Cond. Matter {\bf 19}, 083202 (2007).

\bibitem{freez2} D. Li {\it et al.\/},
Phys. Rev. Lett. {\bf 98}, 190401 (2007).

\bibitem{NielsenChuangBook} M. A. Nielsen and I. L. Chuang, 
{\it Quantum Computations and Quantum
Information} (Cambridge University Press, Cambridge, 2002).

\bibitem{ddJJ1} J. M. Chow {\it et al.\/}, 
Phys. Rev. Lett. {\bf 102}, 090502 (2009); 

%


%
%
%
\bibitem{GerDyb} B. C. Gerstein and C. R. Dybowski, {\it Transient Techniques
in NMR of Solids\/} (Academic Press, Orlando, 1985).

\bibitem{BurumEtal81} D. P. Burum, M. Linder, and R. R. Ernst, J. Mag. Res.
{\bf 43}, 463 (1981).

\bibitem{ShakaEtal88} A. J. Shaka {\it et al.\/}, 
J. Mag. Res. {\bf 80}, 96 (1988).

\bibitem{Haub79} U. Haubenreisser and B. Schnabel, 
J. Mag. Res. {\bf 35}, 175 (1979).
%


\bibitem{decohDuringPulses} See the supplementary material.

\bibitem{ddNV1} F. Jelezko and J. Wrachtrup, J. Phys.: Cond. Matter {\bf 16},
R1089 (2004).

\bibitem{qptNV} M. Howard {\it et al.\/}, 
New J. Phys. {\bf 8}, 33 (2006).

\bibitem{Ref2norm} The 2-norm $||M||_2$ is based on a 
quadratic function of the elements of the $\chi$ and $\chi_0$ matrices;
hence it includes the rms of experimental noise. The fidelity $F$,
for a fixed $\chi_0$, is a linear combination 
of the elements of $\chi$.


\end{thebibliography}
\end{document}